\documentclass[fleqn,usenatbib]{mnras}
\usepackage{newtxtext}
\usepackage[T1]{fontenc}
\usepackage{ae,aecompl}
\usepackage{graphicx}    
\usepackage{amsmath}     
\usepackage{amssymb}     

\usepackage{newtxmath}
\usepackage{multicol}    
\usepackage{bm}          
\usepackage{pdflscape}   
\usepackage{xspace}
\newcommand{\Msun}{\,$M_{\sun}$\xspace}
\newcommand{\Rsun}{\,$R_{\sun}$\xspace}
\newcommand{\Msyr}{\,$M_{\sun}$\,yr$^{-1}$\xspace}
\newcommand{\kms}{\,km\,s$^{-1}$\xspace}
\newcommand{\ergs}{\,erg\,s$^{-1}$\xspace}
\newcommand{\gcm}{\,g\,cm$^{-1}$\xspace}
\newcommand{\cmq}{\,cm$^{-3}$\xspace}
\newcommand{\Ha}{H$\alpha$\xspace}
\newcommand{\Hb}{H$\beta$\xspace}
\newcommand{\HII}{\ion{H}{ii}\xspace}
\newcommand{\HeI}{\ion{He}{i}\xspace}
\newcommand{\HeII}{\ion{He}{ii}\xspace}
\newcommand{\NII}{\ion{N}{ii}\xspace}
\newcommand{\NIII}{\ion{N}{iii}\xspace}
\newcommand{\OI}{\ion{O}{i}\xspace}
\newcommand{\NaI}{\ion{Na}{i}\xspace}

\newcommand{\FeII}{\ion{Fe}{ii}\xspace}
\newcommand{\A}{\,\AA\xspace}
\title[Uncommon SN~2020jfo] 
{Uncommon SN~2020jfo: Ordinary explosion of 8\Msun red supergiant
   with dense wind}
\author[V. P. Utrobin \& N. N. Chugai]{
V. P. Utrobin$^{1,2}$\thanks{E-mail: utrobin@itep.ru}
and
N. N. Chugai$^{2}$\thanks{E-mail: nchugai@inasan.ru}
\\
$^{1}$NRC ``Kurchatov Institute'', acad. Kurchatov Square 1, 123182 Moscow,
      Russia \\
$^{2}$Institute of Astronomy, Russian Academy of Sciences, Pyatnitskaya
      St. 48, 119017 Moscow, Russia
}

\date{Accepted 2023 November 22. Received 2023 November 22; in original form 2023 September 19}

\pubyear{2023}

\begin{document}
\label{firstpage}
\pagerange{\pageref{firstpage}--\pageref{lastpage}}
\maketitle
%
\begin{abstract}
We present the hydrodynamic model of Type IIP SN~2020jfo with the unusually
   short ($\sim$60 days) light curve plateau.
The model suggests the explosion of $\approx$8\Msun red supergiant that ejected
   $\approx$6\Msun with the energy of $\approx$0.8$\times10^{51}$\,erg.
The presupernova wind density turns out highest among known SNe~IIP.
Yet the presupernova was not embedded into a very dense confined circumstellar
   shell that is a feature of some Type IIP supernovae, so 
   the circumstellar interaction in close environment does not contribute
   noticeably to the initial ($\sim$10 days) bolometric luminosity.
Despite uncommon appearance SN~2020jfo turns out similar to SN~1970G
   in the $V$-band light curve, photospheric velocities, and, possibly,
   luminosity as well.
\end{abstract}

\section{Introduction} 
\label{sec:intro}
Type IIP supernova (SN IIP) is an exploding red supergiant with a massive
   hydrogen-rich envelope responsible for the long ($\sim$100 days) light
   curve plateau.
SN~2020jfo in the galaxy M61 received a special attention \citep{Sollerman_2021,
   Teja_2022, Ailawadhi_2023, Kilpatrick_2023} due to its unusually short
   ($\sim$60 days) light curve plateau, which was immediately interpreted as
   a signature of the low ejected mass.
Using the semi-analytic Monte Carlo code to model the bolometric light curve,
   \citet{Sollerman_2021} found that SN~2020jfo has the ejecta mass of
   $\sim$5\Msun.
The hydrodynamic modelling based on the {\sc MESA+STELLA} code package
   suggests a 5\Msun ejecta with the energy of $(2-4)\times10^{50}$\,erg
   interacting with a massive (0.2\Msun) circumstellar (CS) shell in close
   vicinity $r\lesssim6\times10^{14}$\,cm \citep{Teja_2022}.

Although massive CS shell in close vicinity of SN~II ($r< 10^{15}$\,cm)
   generally is not ruled out --- e.g., SN~1998S \citep{Chugai_2001} ---
   there is some doubt that SN~2020jfo harbors so massive confined CS shell.
Indeed, early spectra do not show narrow CS emission lines with broad
   Thomson wings expected in this case, likewise in SN~1998S.

The latter remark is among reasons to revisit the hydrodynamic modelling
   of SN~2020jfo and to present an alternative viewpoint on this highly
   interesting SN~IIP.
Moreover, the study of the CS interaction revealed by optical spectra can
   provide us with the density of the CS matter (CSM) in close vicinity
   and thus independently elucidate the issue of the CS interaction
   in the bolometric luminosity.

In Section~\ref{sec:model} we describe the hydrodynamic model, emphasizing 
   the diagnostic role of the ejecta velocity, and present modelling results.
We then use the observational estimate of a boundary thin shell velocity
   to recover the CSM density including the CS mass within $r<10^{15}$\,cm
    (Section~\ref{sec:wind}).
In Discussion section we consider a case of SN~1970G that turns out
   similar to SN~2020jfo.

Below we use the distance modulus $\mu = 30.81\pm0.20$ ($D = 14.5$\,Mpc)
   and the reddening $E(B-V) = 0.079\pm0.03$\,mag \citep{Kilpatrick_2023},
   and adopt the explosion date of MJD 58973.834 that
   is suggested by the model fit to the rising part of the $r$ light curve.

\section{Hydrodynamic model}
\label{sec:model}
%
\subsection{Model overview}
\label{sec:model-overview}
In a standard approach the SN~IIP hydrodynamic model is constrained by
   the light curve and expansion velocity at the photosphere.
We focus here on the maximum ejecta velocity that is seldom recovered from
   early featureless spectra of SNe~IIP.
Fortunately, for SN~2020jfo we are able to measure the maximum velocity
   based on the broad \HeII 4686\A emission seen on day 2.1
   \citep{Ailawadhi_2023} and day 2.8 \citep{Teja_2022}; the line looks
   similar to the \HeII 4686\A line in SN~2013fs on day 2.4
   \citep{Bullivant_2018}. 

The diagnostic role of the boundary expansion velocity is two-fold.
First, this velocity permits us to constrain CS density and thus to verify
   the models with a massive CS shell for this particular SN.
Second, the maximum velocity provides an additional constraint on presupernova
   (pre-SN) and explosion parameters.
Note that everywhere in the paper ``presupernova'' stands for an exploding
   star and ``progenitor'' stands for a ZAMS star.
The broad \HeII 4686\A emission in the SN~2020jfo spectrum on day 2.1 is
   modelled below (Section \ref{sec:wind}) to derive velocity of the boundary
   thin shell of $16500\pm700$\kms.

We use the radiation hydrodynamics code {\sc CRAB} with the radiation transfer
   in the gray approximation \citep{Utrobin_2004, Utrobin_2007}.
The pre-SN is the hydrostatic non-evolutionary red supergiant (RSG) star.
The term ``non-evolutionary'' means that the stellar structure is
   constructed to reproduce the observed light curve.
There are at least three physical reasons that result in modifying the
   evolutionary pre-SN model: three-dimensional (3D) effects of the
   Rayleigh-Taylor mixing during the SN explosion \citep{Utrobin_2017},
   insufficiently explored effects of a vigorous convection in the RSG
   envelope, and possible binary merger effects \citep{Eldridge_2018}.
The use of the non-evolutionary pre-SN is justified by the modelling of 
   an extended sample of SNe~IIP \citep{Utrobin_2021} including the 3D
   explosions of the evolutionary pre-SN in the case of SN~1999em
   \citep{Utrobin_2017}.

\begin{figure}
   \includegraphics[width=\columnwidth, clip, trim=0 237 54 139]{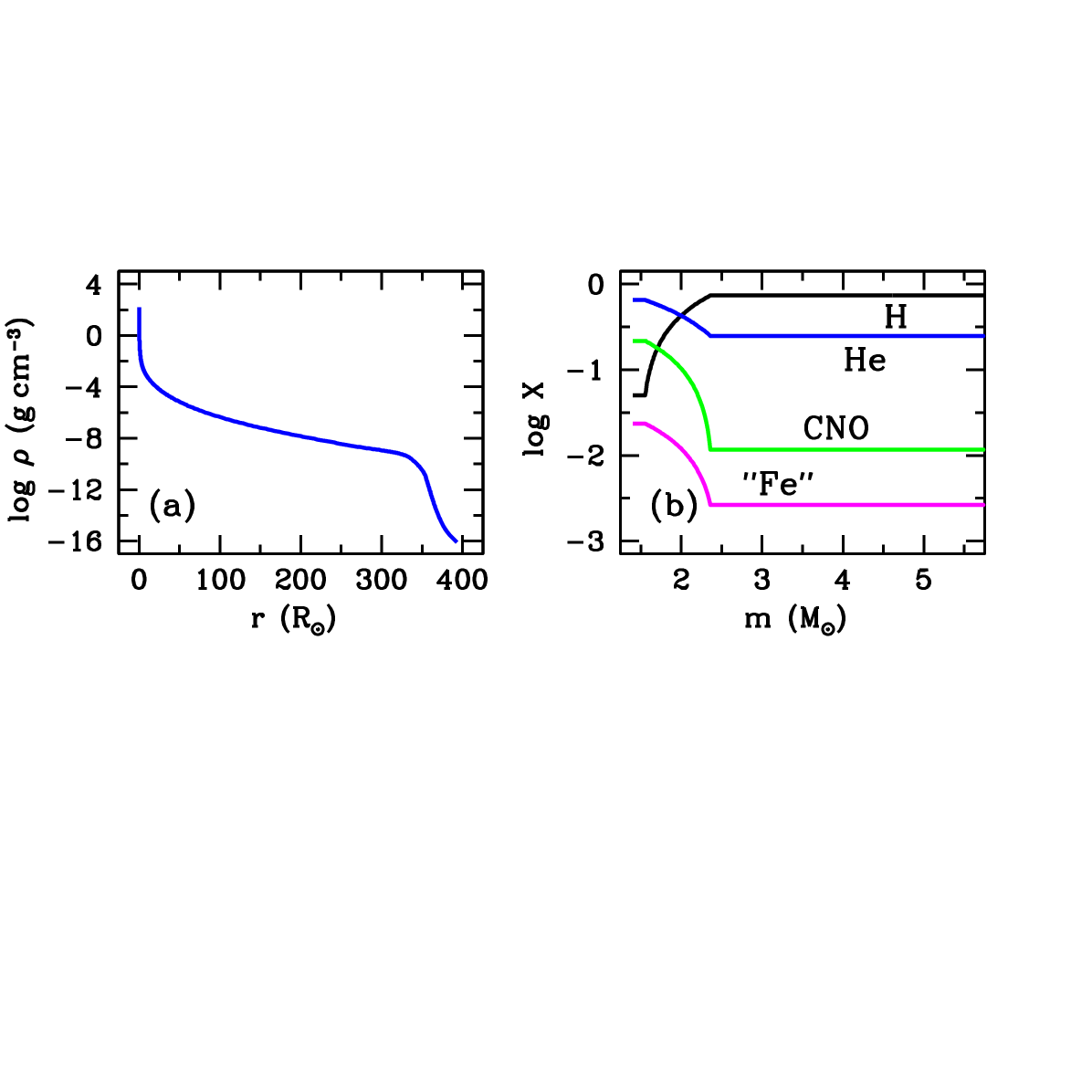}
   \caption{%
   The structure of the pre-SN model.
   Panel (a): the density distribution as a function of radius.
   Panel (b): the chemical composition.
   Mass fraction of hydrogen (\emph{black line\/}), helium
      (\emph{blue line\/}), CNO elements (\emph{green line\/}),
      and Fe-peak elements excluding radioactive $^{56}$Ni
      (\emph{magenta line\/}) in the ejected envelope.
   The central core of 1.4\Msun is omitted.
   }
   \label{fig:presn}
\end{figure}
The explosion is initiated by a supersonic piston applied to the stellar
   envelope at the boundary with the 1.4\Msun collapsing core.
The description of the light curve and velocities at the photosphere,
   including the outermost expansion velocity, requires a fine tuning of
   the density distribution and chemical composition.
The optimal pre-SN model has a smooth density and composition gradients
   at the metals/He and He/H interfaces (Fig.~\ref{fig:presn}), which mimics
   the mixing outcome in 3D hydrodynamic simulations of SN~IIP explosion
   \citep{Utrobin_2017}.

The mixing of radioactive $^{56}$Ni affects the light curve as well.
Increasing the 56Ni mixing in velocity space makes the plateau slightly
   brighter and its duration shorter \citep{Utrobin_2007}.
A more extended $^{56}$Ni mixing, when the escape of the gamma rays becomes
   essential, also affects the rate of the radiactive tail decay resulting in
   a more rapid decay compared to the $^{56}$Co decay rate.
The outer velocity of the $^{56}$Ni ejecta is well constrained by the width
   of \HeI emission lines that are highly sensitive to the $^{56}$Ni
   distribution \citep{Lucy_1991,Utrobin_1996}.
Unfortunately, \HeI 5876\A line is blended with \NaI doublet. 
Less prominent but well observed \HeI 7065\A emission in spectra taken
   at NOT/ALFOSC on days 216 and 280 (WISeREP archive) has a blue width
   at zero intensity of 1300\kms and 1600\kms, respectively.
These velocities suggest the outer boundary of $^{56}$Ni ejecta at about
   1500\kms that is consistent with 1600\kms adopted in the hydrodynamic
   model.

\subsection{Results}
\label{sec:model-results}
%
\begin{figure}
   \includegraphics[width=\columnwidth, clip, trim=0 238  54 133]{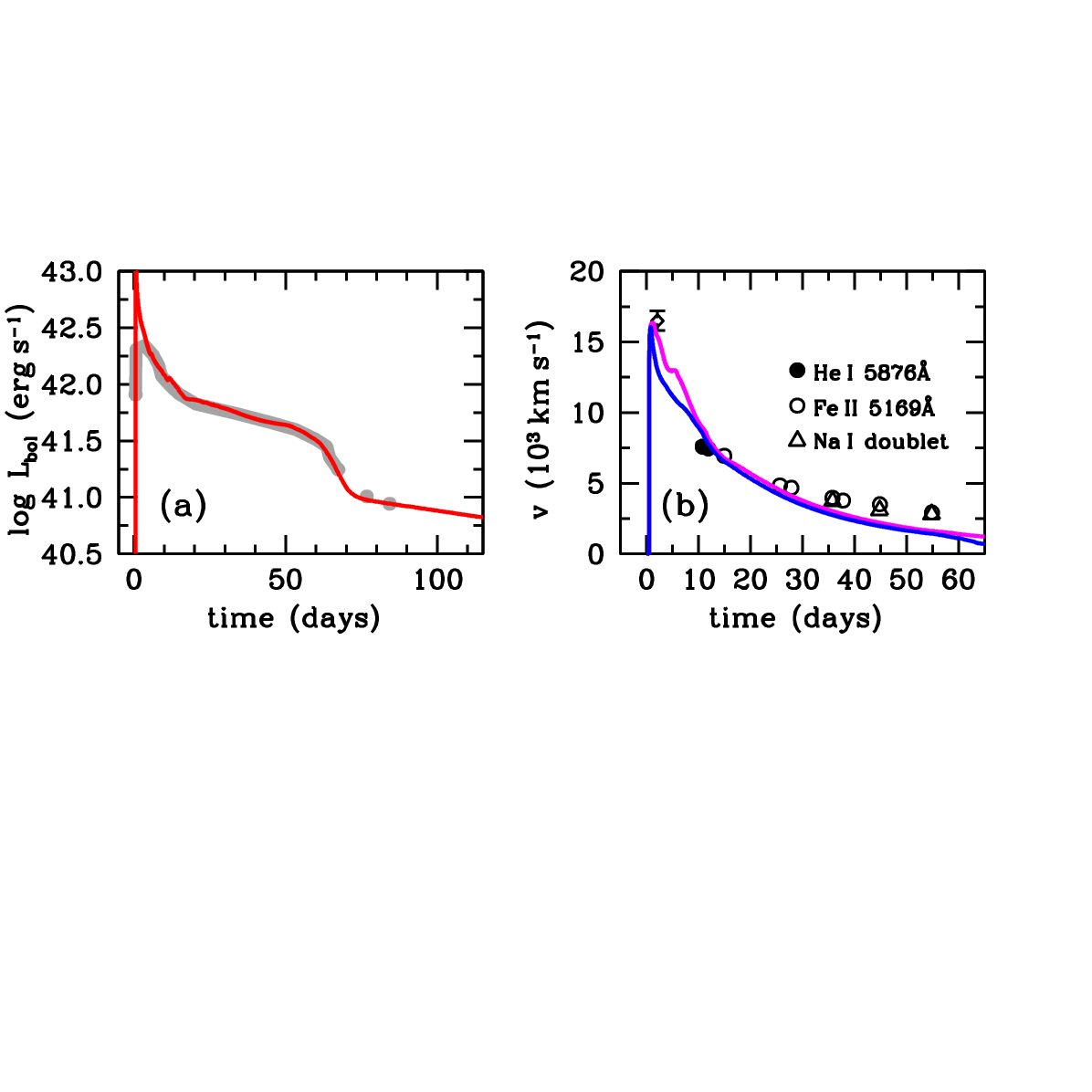}
   \caption{%
   The bolometric light curve and the evolution of photospheric velocity.
   Panel (a): the model light curve (\emph{red line\/}) overlaid on
      the bolometric data (\emph{gray circles\/}) \citep{Kilpatrick_2023}.
   Panel (b): the evolution of model velocity defined by the level
      $\tau_{eff} = 2/3$ (\emph{blue line\/}) and $\tau_\mathrm{Thomson} = 1$
      (\emph{magenta line\/}) is compared with the photospheric velocities
      estimated from the absorption minimum of \FeII 5169\A \citep{Teja_2022}
      along with our estimates from the \HeI 5876\A line absorption and 
      the profile of \NaI doublet.
   Note that the maximum ejecta velocity of the hydrodynamic model fits 
      the thin shell velocity on day 2.1 recovered from \HeII 4686\A line
      (\emph{diamond\/}).
   }
   \label{fig:lcv}
\end{figure}
The optimal hydrodynamic model is specified by the ejecta mass
   $M_{ej} = 6.2$\Msun, the explosion energy $E = 0.756\times10^{51}$\,erg,
   and the pre-SN radius $R_0 = 400$\Rsun.
The $^{56}$Ni mass directly recovered from the radioactive tail is 0.013\Msun.
The uncertainty in the derived SN parameters can be estimated by a variation
   of the model parameters around the optimal model.
The uncertainties of the distance and the reddening (see Section~\ref{sec:intro})
   imply the 20 per cent uncertainty in the bolometric luminosity. 
The scatter in the plot of the photospheric velocity versus time
   (Fig.~\ref{fig:lcv}b) suggests the uncertainty of 7 per cent
   in the photospheric velocity. 
We estimate the maximal uncertainty of the plateau length as 2\,days, i.e.
   3 per cent of the plateau duration. 
With these uncertainties of observables, we find the errors of
   $\pm100$\Rsun for the initial radius, $\pm0.52$\Msun for the ejecta
   mass, $\pm0.233\times10^{51}$\,erg for the explosion energy, and
   $\pm0.0026$\Msun for the total $^{56}$Ni mass.

\begin{figure}
   \includegraphics[width=\columnwidth, clip, trim=0 239 54 139]{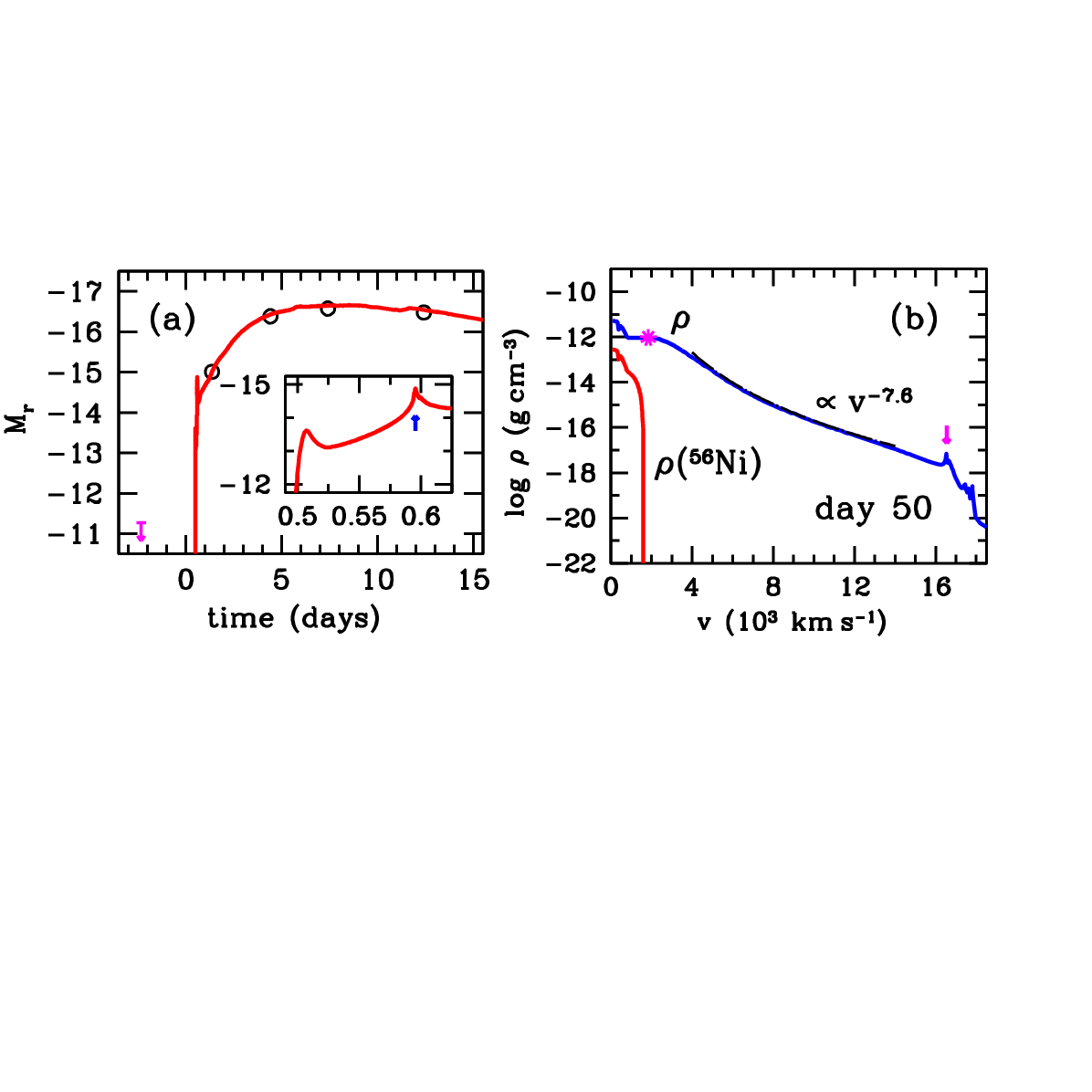}
   \caption{
   Panel (a): Rising part of the model light curve in $r$-band overplotted
      on the observational data taken by the Zwicky Transient Facility.
   The magenta arrow shows the observational upper limit reported by
      \citep{Kilpatrick_2023} about 4.2 days prior to SN discovery.
   Inset shows the fine structure of the narrow peak related to shock breakout.
   The second fine-structure peak indicated by blue arrow corresponds to
      the formation of the thin boundary shell marked by arrow
      on the right panel.
   Panel (b): The density (\emph{blue line\/}) and $^{56}$Ni (\emph{red line\/})
      distributions vs. velocity in the ejecta on day 50; magenta star indicates
      the photosphere location, while magenta arrow shows the boundary thin shell.
   \emph{Dash-dotted line\/} is the power law $\rho \propto v^{-7.6}$.
   }
\label{fig:rise}
\end{figure}

The optimal model provides a good fit to the bolometric light curve along with
   a reasonable description of velocity at the photosphere (Fig.~\ref{fig:lcv}).
The model luminosity peak related to the shock breakout is more luminous
   compared to the observational data, which presumably is caused by
   the missing the far-UV flux in the recovered observational bolometric flux.
The density and $^{56}$Ni distributions in the freely expanding ejecta
   on day 50 are shown in Fig.~\ref{fig:rise}b.
It should be emphasized that the CS interaction is not included in the light
   curve modelling.
This is justified by the moderate density of the CSM suggested by the analysis
   of CS interaction effects in optical spectra (Section~\ref{sec:wind}).

The rising part of the $r$-band light curve after the shock breakout is
   well reproduced (Fig.~\ref{fig:rise}a) that permits us to reliably fix
   the explosion moment.
The initial narrow peak of the model has double-peak structure
   (Fig.~\ref{fig:rise}a, inset), the phenomenon described already for
   SN~2017gmr model \citep{Utrobin_2021}.
Note that the second fine structure peak signals on the formation of the
   boundary shell by the radiation flux at the shock breakout phase.
This shell is seen in the density distribution of freely expanding ejecta 
   (Fig.~\ref{fig:rise}b).
The verification of the double peak structure of the shock breakout peak
   is a challenging task for the early SNe~IIP photometry.

At the late photospheric epoch (>\,40\,days) the model underproduces
   the photospheric velocity compared to the observed one inferred from
   \FeII and \NaI lines.
This mismatch is characteristic of some SNe~IIP, particularly, SN~1999em
   \citep{Utrobin_2017}.
It is noteworthy that this disparity is characteristic of both kind of models:
   the 3D explosion of the evolutionary pre-SN and the 1D explosion of
   the optimal non-evolutionary pre-SN \citep{Utrobin_2017}.
The problem probably reflects features of the pre-SN structure or/and
   the explosion physics missing in our model.
It is interesting that some hydrodynamic models of SN~2020jfo are able
   to reproduce the late photospheric velocities \citep{Teja_2022},
   but at the price of the mismatch between the model and observed luminosities
   at the photospheric epoch.

\section{Presupernova wind}
\label{sec:wind}
%
\subsection{Optical markers of ejecta deceleration}
\label{sec:wind-markers}
A key structure element to our picture of the ejecta/wind interaction is
   the cold dense shell (CDS) that forms between the forward and reverse shocks.
The latter is usually radiative, so the low mass of the boundary thin shell
   that forms by the shock breakout ($\sim$10$^{-4}$\Msun) in a couple of days
   can mount up to $\sim$10$^{-3}$\Msun due to the CS interaction with 
   a sufficiently dense wind ($w = \dot{M}/u_w \sim 10^{15}$\gcm).
At this early stage the CDS turns out optically thick in the continuum,
   which explains featureless spectrum during the first 3 days
   \citep{Ailawadhi_2023, Teja_2022}.
The two kind of optical markers of the CS interaction in SN~2020jfo are
   closely related to the CDS. 

The first marker is a low-contrast broad emission \HeII 4686\A
   in the spectrum on days 2.1 and 2.8 \citep{Teja_2022,Ailawadhi_2023}
   that is almost exact replica of the \HeII line in SN~2013fs on day 2.4
   \citep{Bullivant_2018}.
This line with a strongly blueshifted profile is emitted by a narrow shell
   that is opaque at line frequencies and resides on the top of the photosphere
   coincident with the CDS \citep{Chugai_2020}.
To infer a reliable velocity of the line-emitting shell, we consider a simple
   model and compute the line profile using the Monte-Carlo technique.
The model suggests that the \HeII line is emitted by a narrow
   ($\Delta r/r = 0.08$) shell with the Sobolev optical depth
   $\gg 1$; the line-emitting shell is attached to the CDS with adopted
   albedo of 0.3.
Line photons can be scattered off thermal electrons with the temperature of
   13000\,K in the wind with the Thomson optical depth of 0.3.
The computed line overlaid on the observed spectrum on day 2.1
   (Fig.~\ref{fig:wind}a, bottom inset) implies the shell velocity
   of $16500\pm700$\kms.
The observed flux excess in blue and red wings are due to unaccounted emission
   of \NIII doublet and \Hb, respectively.

The second marker is a high-velocity narrow absorption (HVNA) in the \Ha blue
   absorption wing at the age of 45 and 55 days \citep{Teja_2022}. 
The HVNA in \Ha and \Hb have been discovered and dubbed ``notch'' in SN~1999em
   \citep{Leonard_2002}.
Subsequently HVNA has been attributed to the absorption of the photospheric
   radiation in the CDS between the forward and reverse shocks
   \citep{Chugai_2007}.
The observed HVNA velocity in SN~2020jfo is lower than the measured CDS velocity
   by $\approx$210\kms; the correction is related to the intensity integration
   over the visible photosphere.
The resulting measured CDS velocity is 9570\kms and 9435\kms on days 45 and 55,
   respectively.
The HVNA in \Hb cannot be identified reliably because of irregular wiggles
   in the blue wing of the broad absorption trough.
At least the spectrum on day 55 shows in the \Hb blue wing small ``notch''
   at the correct velocity of $\approx -9200$\kms.

With the inferred velocities from the \HeII line and HVNA, the observed CDS
   velocity is fixed at three epochs: 2.1 days, 45 days, and 55 days
   (Fig.~\ref{fig:wind}a). 

\subsection{Density of presupernova wind}
\label{sec:wind-density}
%
\begin{figure}
   \includegraphics[width=\columnwidth, clip, trim=0 132  10 133]{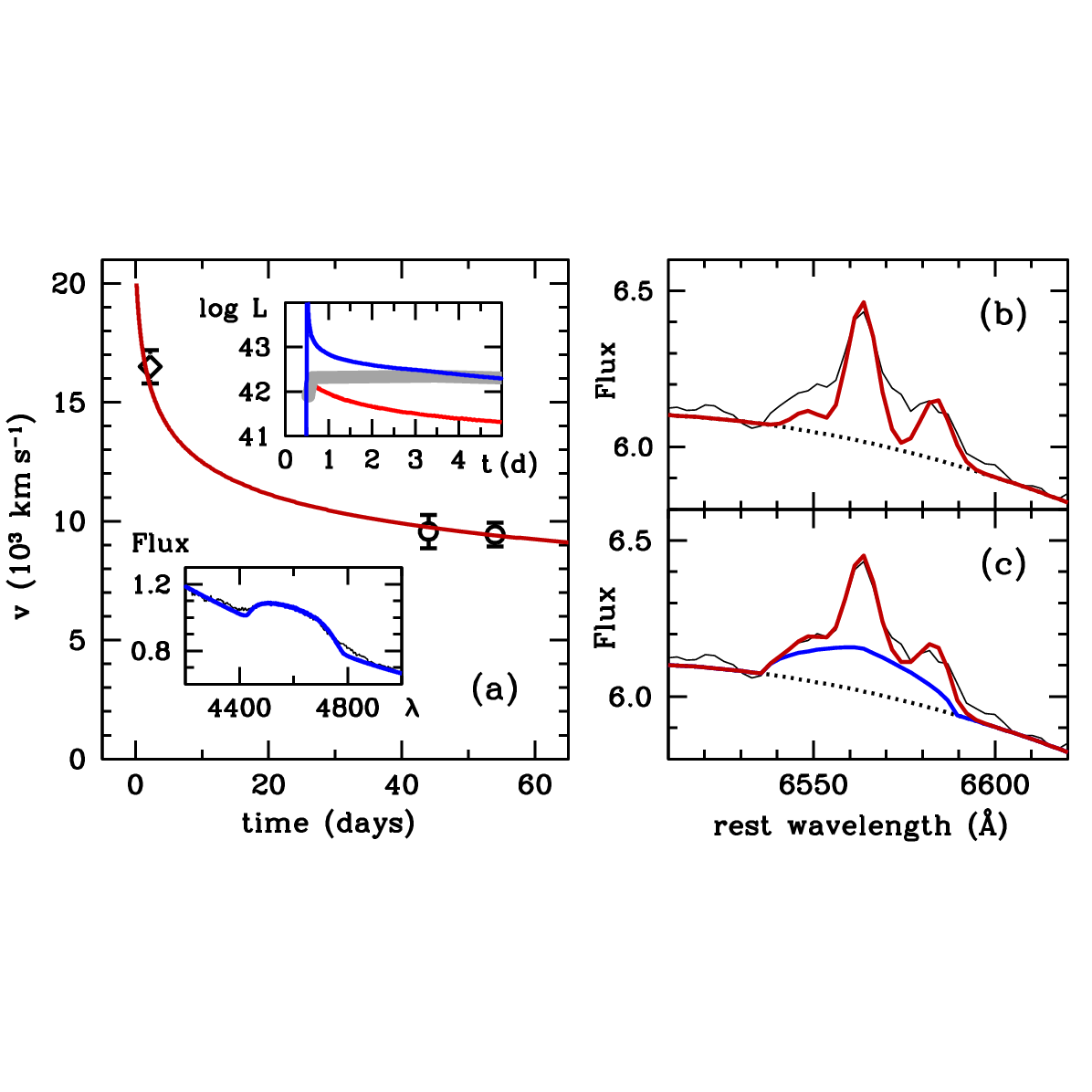}
   \caption{
   Panel (a): Model CDS velocity (\emph{red line\/}) overlaid on the
      observational CDS velocity recovered from \HeII 4686\A (\emph{diamond\/})
      and \Ha HVNA (\emph{circles\/}).
   Bottom inset shows the model profile of \HeII 4686\A line (\emph{blue line\/})
      overlaid on the observed spectrum (\emph{thin black line\/}) on day 2.1
      \citep{Ailawadhi_2023}.
   Top inset shows the optical bolometric luminosity powered solely
      by CS interaction (\emph{red line\/}) compared to the light curve
      (\emph{blue line\/}) produced by the hydrodynamic model without the CS
      interaction and to the bolometric data (\emph{gray band\/})
      \citep{Kilpatrick_2023}.
   Panel (b): the synthetic spectrum (\emph{red line\/}) composed by the \Ha and
      [\NII] lines from \HII region of the host galaxy without the contribution
      of \Ha from CS wind is compared with the observed one (\emph{thin black
      line\/}) on day 2.1.
   \emph{Dotted line\/} is the approximation of the underlying broad-band
      spectrum.
   Panel (c): the same as Panel (b) but with the contribution of \Ha from CSM
      with the maximum velocity of 1200\kms (\emph{blue line\/}).
   }
   \label{fig:wind}
\end{figure}
The CS interaction model  \citep{Chugai_2018} is based on the thin shell
   approximation \citep{Chevalier_1982}.
The evolution of the CDS velocity depends on the density distribution of
   the ejecta $\rho(v)$ and the CS density distribution $\rho(r)$.
The density distribution in the freely expanding ejecta is approximated by
   the analytic expression $\rho = \rho_0(t_0/t)^3/[1 + (v/v_0)^{7.6}]$
   suggested by the hydrodynamic model (Fig.~\ref{fig:rise}b);
   $\rho_0$ and $v_0$ are defined by the ejecta mass $M$ and the kinetic
   energy $E$.
The CS wind is assumed to be steady with the density distribution
   $\rho \propto r^{-2}$.

The optimal CS interaction model describes the CDS evolution for the wind
   density parameter $w = \dot{M}/u = 2.2\times10^{15}$\gcm that corresponds
   to the mass-loss rate $\dot{M} = 5.2\times10^{-5}u_{15}$\Msyr,  
   where $u_{15}$ is the wind speed in units of 15\kms.
The latter value is adopted following the wind speed of Betelgeuse
   \citep{Smith_2009}.
The optical luminosity powered by the CS interaction is dominated by
   the forward shock.
This luminosity is significant during the first day, but still substantially
   weaker compared to the SN radiation in the hydrodynamic model without
   the CS interaction (Fig.~\ref{fig:wind}a, top inset).
This justifies the omission of the CSM in the hydrodynamic model.

The emission measure of the preshock wind on day 2.1 is EM = $3\times10^{62}$\cmq.
Assuming that on day 2.1 the photosphere resides at the CDS
   ($r = 3.2\times10^{14}$\,cm), from the observed luminosity of
   $\approx 2\times10^{42}$\ergs one finds the effective temperature of
   $\approx 1.3\times10^4$\,K.
Adopting the same kinetic temperature, one obtains the \Ha effective
   recombination coefficient $\alpha_{32} = 9.1\times10^{-14}$\cmq\,s$^{-1}$
   \citep{Osterbrock_2006}.
The recovered wind density thus suggests the \Ha luminosity of fully ionized
   wind on day 2.1 of $\approx 8\times10^{37}$\ergs.

We examine the available spectrum on day 2.1 and find at the \Ha position
   a blend of narrow \Ha and [\NII] 6548, 6583\A apparently from a superimposed
   \HII region in the host galaxy (Fig.~\ref{fig:wind}b).
However, an attempt to attribute a broad emission feature (full width at zero
   intensity of 2400\kms) to the narrow \Ha combined with \NII lines obviously
   fails: an additional broad emission in the \Ha band is certainly needed.
The synthetic spectrum composed by the ``hand-made'' \Ha emission in the range
   $|v_r| \lesssim 1200$\kms with the equivalent width of $0.82\pm0.05$\A
   and by the emission lines of \HII region provides an acceptable description
   of the observed emission blend (Fig.~\ref{fig:wind}c).
We attribute the origin of the ``broad'' \Ha emission to the preshock wind
   accelerated up to $\approx$1200\kms by the SN radiation.
The preshock wind velocities of $\sim$10$^3$\kms at about 2 days after
   the shock breakout in SN~IIP is an expected outcome of the radiative
   acceleration \citep{Dessart_2017}.

For the black body continuum with the temperature of 13000\,K on day 2.1 and
   the bolometric luminosity of $\approx$2$\times10^{42}$\ergs
   the equivalent width of 0.82\A implies the \Ha luminosity of
   $\approx$(7.3$\pm0.5)\times10^{37}$\ergs, which is in a reasonable
   agreement with the \Ha luminosity of the wind ($\approx$8$\times10^{37}$\ergs).

The CSM mass in the region $r < 10^{15}$\,cm (a typical scale for the confined
   dense CS shell) turns out $10^{-3}$\Msun.
This is by two orders below than needed ($\sim$0.2\Msun) to account for the
   initial luminosity peak by means of the CS interaction \citep{Teja_2022}.
To summarize, the inferred wind density implies that SN~2020jfo exploded in
   a dense CS wind that however is not dense enough to affect the SN bolometric
   luminosity.

\begin{figure}
   \includegraphics[width=\columnwidth, clip, trim=8 23 29 28]{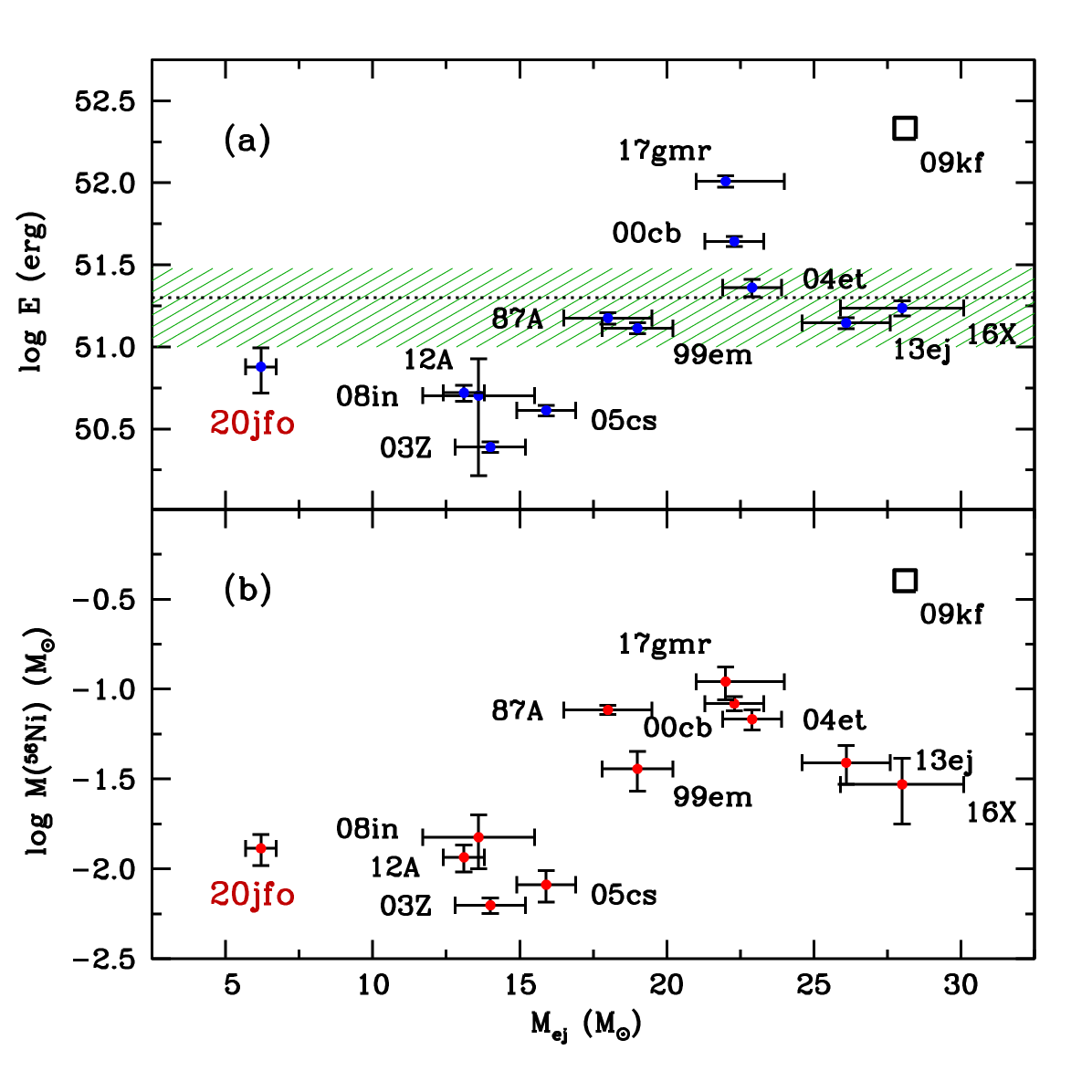}
   \caption{
   Explosion energy (Panel (a)) and $^{56}$Ni mass (Panel (b))
      versus ejecta mass for SN~2020jfo and twelve other
      core-collapse SNe studied using the uniform approach \citep{Utrobin_2021}.
   Dotted line in Panel (a) is the upper limit of the explosion energy of
      $2\times10^{51}$\,erg for the neutrino-driven mechanism \citep{Janka_2017}
      with the uncertainty of about $\pm 10^{51}$\,erg\protect\footnotemark\
      shown by the shaded \emph{green band}.
   }
   \label{fig:ennims}
\end{figure}
\footnotetext{H.-Th.~Janka, private communication.}  
%

\section{Discussion}
\label{sec:disc}
%
\begin{table}
\centering
\caption[]{Hydrodynamic models of Type IIP supernovae.}
\label{tab:sumtab}
\begin{tabular}{@{ } l  c  c @{ } c @{ } c @{ } c  c @{ }}
\hline
\noalign{\smallskip}
 SN & $R_0$ & $M_{ej}$ & $E$ & $M_{\mathrm{Ni}}$ 
       & $v_{\mathrm{Ni}}^{max}$ & $v_{\mathrm{H}}^{min}$ \\
       & (\Rsun) & (\Msun) & ($10^{51}$\,erg) & ($10^{-2}$\Msun)
       & \multicolumn{2}{c}{(km\,s$^{-1}$)}\\
\noalign{\smallskip}
\hline
\noalign{\smallskip}
 1987A  &  35  & 18   & 1.5    & 7.65 &  3000 & 600 \\
1999em  & 500  & 19   & 1.3    & 3.6  &  660  & 700 \\
2000cb  &  35  & 22.3 & 4.4    & 8.3  &  8400 & 440 \\
 2003Z  & 230  & 14   & 0.245  & 0.63 &  535  & 360 \\
2004et  & 1500 & 22.9 & 2.3    & 6.8  &  1000 & 300 \\
2005cs  & 600  & 15.9 & 0.41   & 0.82 &  610  & 300 \\
2008in  & 570  & 13.6 & 0.505  & 1.5  &  770  & 490 \\
2009kf  & 2000 & 28.1 & 21.5   & 40.0 &  7700 & 410 \\
2012A   &  715 & 13.1 & 0.525  & 1.16 &  710  & 400 \\
2013ej  & 1500 & 26.1 & 1.4    & 3.9  &  6500 & 800 \\
 2016X  &  436 & 28.0 & 1.73   & 2.95 &  4000 & 760 \\
2017gmr &  525 & 22.0 & 10.2   & 11.0 &  3300 & 640 \\
2020jfo &  400 &  6.2 & 0.756  & 1.3  &  1600 & 190 \\
\noalign{\smallskip}
\hline
\end{tabular}
\end{table}
The hydrodynamic modelling of SN~2020jfo results in the picture of the explosion
   of $\approx$8\Msun red supergiant with the energy of
   $\approx$0.76$\times10^{51}$\,erg and the ejection of an envelope of
   $\approx$6\Msun and a mass of radioactive $^{56}$Ni of 0.013\Msun.
The explosion energy and the ejected $^{56}$Ni mass are consistent with those
   of a neutrino-driven explosion.
While the ejecta mass is comparable to that of the model of \citet{Teja_2022},
   our energy is twice as large and the pre-SN radius of 400\Rsun is
   significantly smaller compared to that of 680\Rsun in \citet{Teja_2022}.
In the absence of the information on their model photospheric velocity
   at the early stage ($t < 15$\,days) we do not consider these differences
   as meaningful.

With parameters of other dozen SNe~IIP (Table~\ref{tab:sumtab}) recovered via
   the uniform approach \citep{Utrobin_2021}, on the diagrams $E$ vs. $M_{ej}$
   and $M(^{56}\mathrm{Ni}$) vs. $M_{ej}$ (Fig.~\ref{fig:ennims}) the SN~2020jfo
   location is notable: it has the lowest ejecta mass, the ordinary energy
   for SNe~IIP, and the amount of $^{56}$Ni comparable to that of SNe~IIP
   with a low ejected mass.
All in all, given a large scatter of SNe~IIP on the diagrams, the location of
   SN~2020jfo does not contradict to the conjecture of the low mass
   ($\sim$12\Msun) progenitor.

\begin{figure}
   \includegraphics[width=\columnwidth, clip, trim=18 114 28 116]{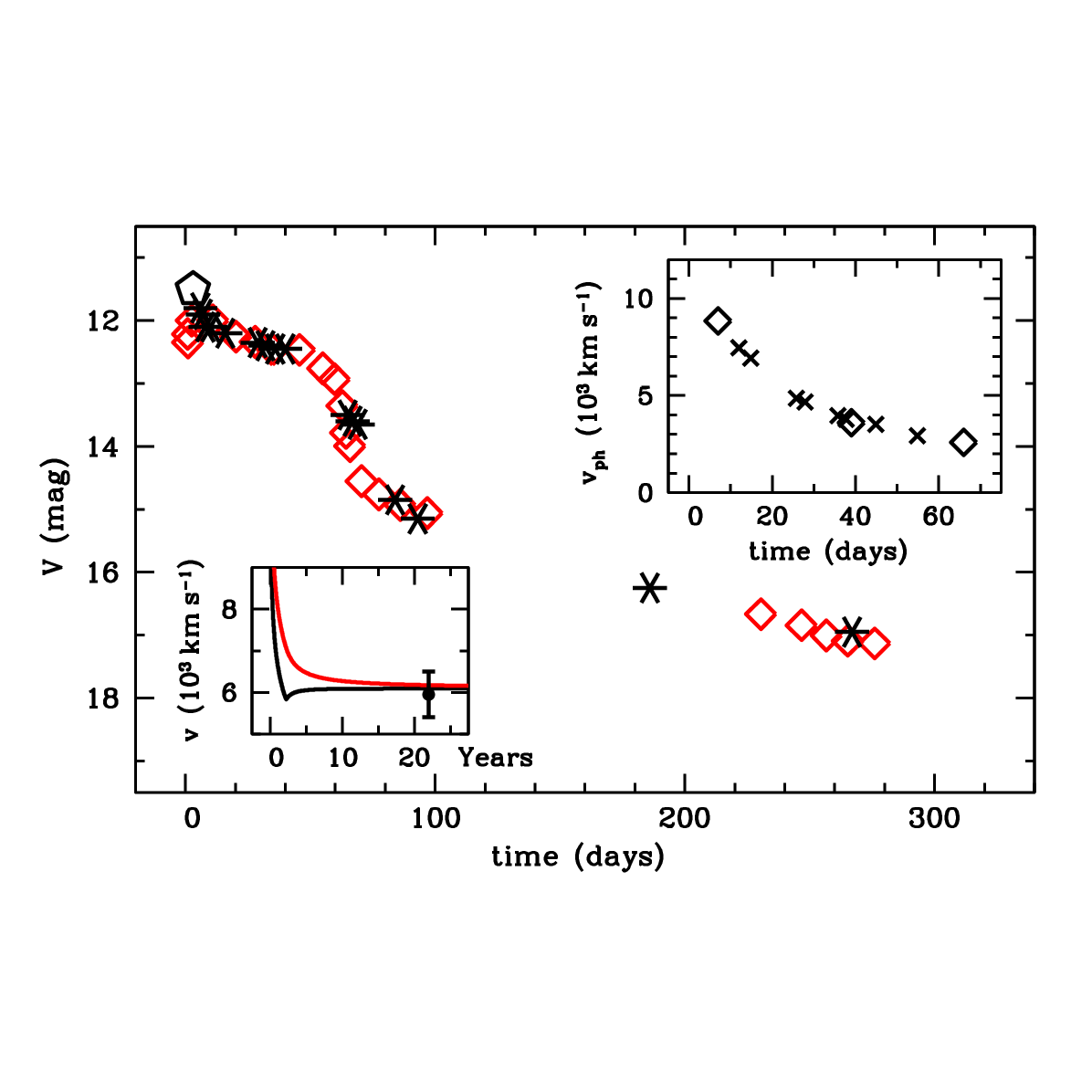}
   \caption{
   Light curves in $V$-band of SN~2020jfo
     \citep[][\emph{red diamonds\/}]{Ailawadhi_2023} and of SN~1970G
     (\emph{asteriscs\/}).
   The discovery $V$ magnitude of SN~1970G is shown by \emph{pentagon\/} symbol.
   Top inset shows the photospheric velocity of SN~1970G (\emph{diamonds\/})
      compared to the photospheric velocity of SN~2020jfo (\emph{crosses\/}).
   Bottom inset shows the evolution of model velocities of the CDS for SN~2020jfo
      (\emph{black line\/}) and the boundary velocity of unshocked ejecta
      (\emph{red line\/}).
   The point is the maximum expansion velocity of SN~1970G recovered from
      emission lines of [\OI]\,6300, 6364\A and \Ha.
   }
   \label{fig:70g}
\end{figure}
Despite, at first glance, an unique SN~2020jfo light curve, we find yet another
   SN~IIP with the similar short plateau --- SN~1970G in M101 \citep{Barbon_1973}.
The resemblance between SN~1970G and SN~2020jfo is demonstrated by the similarity
   of their $V$ light curves from \citet{Barbon_1973} and \citet{Ailawadhi_2023},
   respectively (Fig.~\ref{fig:70g}).
The adopted explosion date for SN~1970G is July 27, three days before its
   discovery on July 30 \citep{Lovas_1970}.
The photospheric velocity on day 7 \citep{Barbon_1973} as well as on days
   39 and 66 \citep{Chugai_1982} are consistent to the velocities of SN~2020jfo
   \citep{Teja_2022} (Fig.~\ref{fig:70g}, top inset).
Moreover, the similarity concerns also the pre-SN wind: the predicted
   deceleration of the outermost ejecta of SN~2020jfo at the age of 22\,yr
   is consistent with the observed maximum expansion velocity of SN~1970G
   (Fig.~\ref{fig:70g}, bottom inset) recovered from [\OI]\,6300,6364\A and
   \Ha at this age \citep{Fesen_1993}.
The mentioned similarities suggest similarity of their ejecta mass, energy,
   and the wind density.

In order to superimpose both light curves, we shift the SN~2020jfo light curve
   down by $\Delta V = 2.4$\,mag that primarily could be attributed to the
   differences in their distances and extinctions.
If we use the distance to M101 of 6.85\,Mpc \citep{Riess_2022} and 14.5\,Mpc
   to M61 \citep{Ailawadhi_2023},  $A_V = 0.13$\,mag for SN~1970G, the same as
   for SN~2023ixf in M101 \citep{Jacobson-Galan_2023}, and $A_V = 0.45$\,mag
   for SN~2020jfo --- the average between adopted values of \cite{Ailawadhi_2023}
   and \cite{Kilpatrick_2023} --- we find that SN~1970G turns out brighter by
   0.45\,mag in absolute magnitude.
Although plausible, the recovered difference is not reliable because the distance
   and extinction for SN~2020jfo are not free from errors.
We therefore conservatively consider the bolometric luminosity of both
   supernovae comparable.

Yet, we note that the initial behavior of the $V$-band flux is somewhat different:
   the $V$ light curve of SN~1970G shows a sharp initial maximum, whereas
   SN~2020jfo does not.
This might stem from the larger pre-SN radius of SN~1970G at the explosion.
In this respect one has to take into account that RSGs are generally pulsating
   stars, so two SNe with similar mass, explosion energy, and the average radii
   may show somewhat different initial luminosity peak, if they explode
   at different pulsation phases.

The derived 6.2\Msun ejecta mass combined with the 1.4\Msun collapsing core
   implies that the SN~2020jfo pre-SN at the explosion was the 7.6\Msun RSG.
This should be considered as a lower limit mass for the ZAMS progenitor.
With the recovered mass-loss rate of $\approx$5$\times10^{-5}u_{15}$\Msyr
   the RSG progenitor on the time scale of $\sim$10$^5$\,yr was able
   to lose several solar masses, so the ZAMS progenitor might well be
   an $\approx$12\Msun star preferred by recent studies \citep{Sollerman_2021,
   Teja_2022, Ailawadhi_2023}.
Yet we emphasize that we are not able to recover the progenitor (ZAMS) mass
   from our results and bound ourselves with the inferred pre-SN mass of 8\Msun.

Remarkably, the wind density of SN~2020jfo turns out maximum among known SNe~IIP.
Indeed, the mass-loss rate of four SNe~IIP (SN~1999em, SN~2002hh, SN~2004dj,
   and SN~2004et) recovered from the radio flux evolution \citep{Chevalier_2006}
   lie in the range of $(0.4-1.5)\times10^{-5}u_{15}$\Msyr with the maximum
   value by a factor of three lower compared to the wind of SN~2020jfo.
Although high, this mass-loss rate is within the range suggested by the classic
   value based on the momentum conservation for the wind with the optical depth
   $\tau \sim 1$, viz. $\dot{M} = L/(uc) = 7\times10^{-5}$\Msyr, where we use
   $u = 15$\kms and the 12\Msun progenitor luminosity of $2\times10^{38}$\ergs
   \citep{Meynet_2015}.
Noteworthy, despite of the high wind density SN~2020jfo does not show strong
   emission lines in early spectrum on day 2.1 usually considered as a signature
   of a dense confined ($r<10^{15}$\,cm) CS shell likewise in SN~2013fs
   \citep{Yaron_2017}.

While the 8\Msun pre-SN argued in this paper is consistent with the
   progenitor mass inferred from the archive photometry data
   \citep{Kilpatrick_2023}, one cannot rule out that the ZAMS star has been
   essentially more massive, say, about 12\Msun.
Apart from the mass reduction via the RSG wind discussed above, one has to
   admit an interesting possibility related to the progenitor evolution in
   a binary system.
The binary nature of the majority of massive stars suggests at least two
   scenarios for the significant loss of the progenitor mass:
   (i) the non-conservative mass transfer to the less massive component and
   (ii) the merger process.
The population synthesis simulations \citep{Eldridge_2018} show that
   SNe~IIb with the hydrogen envelope mass less than 1\Msun are far more
   likely final outcome for the massive binary evolution than SNe~IIP
   with a (several)$\times$\Msun hydrogen envelope like SN~2020jfo. 
This could account for a scarcity of SNe~IIP with short plateau.

\section{Conclusions}
\label{sec:concl}
We bind ourselves with the emphasis on the four major results:
\begin{itemize}
\item The hydrodynamic modelling of Type IIP supernova SN~2020jfo with short
   plateau suggests the explosion of $\approx$8\Msun RSG with the radius of
   $\approx$400\Rsun that ejects $\approx$6\Msun with the energy of
   $\approx$0.8$\times10^{51}$\,erg.
\item The found pre-SN wind density is highest among known SNe~IIP.
\item The recovered wind density is insufficient for the significant
   contribution into the bolometric light curve.
\item We find that SN~1970G is similar to SN~2020jfo in the $V$ light curve,
   photospheric velocities, and, possibly, the luminosity as well.
\end{itemize}

\section*{Acknowledgements}
The reported study was partially funded by RFBR and DFG, project number
   21-52-12032.
VPU is partially supported by Russian Scientific Foundation grant
   19-12-00229.

\section*{Data Availability}
The data underlying this article will be shared on reasonable request to
   the corresponding author.


\bsp	
\label{lastpage}
\end{document}